\documentclass[3p,times]{elsarticle}

\usepackage{ecrc}

\volume{00}

\firstpage{1}

\journalname{Chaos, Solitons and Fractals}

\runauth{G. Wilk and Z. W\l odarczyk}

\jid{ChSF}

\jnltitlelogo{Chaos Solitons Fractals}

\CopyrightLine{2011}{Published by Elsevier Ltd.}

\usepackage{amssymb}

\usepackage[figuresright]{rotating}

\begin{document}

\begin{frontmatter}



\dochead{}

\title{Quasi-power laws in multiparticle production processes\tnoteref{label1}}
\tnotetext[label1]{Invited talk presented by G.Wilk at
SigmaPhi2014 conference at Rhodes, Greece.}


\author[NCNR]{Grzegorz Wilk\corref{cor1}}
\ead{grzegorz.wilk@fuw.edu.pl}
\author[UJK]{Zbigniew W\l odarczyk}
\ead{zbigniew.wlodarczyk@ujk.edu.pl}

\cortext[cor1]{Corresponding author}

\address[NCNR]{National Centre for Nuclear Research,~Department of
Fundamental Research, Ho\.z a 69, 00-681; Warsaw, Poland }

\address[UJK]{Institute of Physics, Jan Kochanowski University,
\'Swi\c{e}tokrzyska 15; 25-406 Kielce, Poland}

\begin{abstract}
We review the ubiquitous presence in multiparticle production
processes of quasi-power law distributions (i.e., distributions
following pure power laws for large values of the argument but
remaining finite, usually exponential, for small values). Special
emphasis is placed on the conjecture that this reflects the
presence in the produced hadronic systems of some intrinsic
fluctuations. If described by parameter q they form, together with
the scale parameter $T$ ("temperature"), basis of Tsallis
distribution, ${f(X)\sim [1 - (1-q)X/T]^{1/(1-q)}}$ , frequently
used to describe the relevant distributions (the X being usually a
transverse momentum). We discuss the origin of such quasi-power
law behavior based on our experience with the description of
multiparticle production processes. In particular, we discuss
Tsallis distribution with complex nonextensivity parameter q and
argue that it is needed to describe log-oscillations as apparently
observed in recent data on large momentum distributions in very
high energy p-p collisions.
\end{abstract}


\end{frontmatter}



\section{Introduction}
\label{sec:I}

\begin{figure}[h]
\begin{center}
\includegraphics[width=9.5cm]{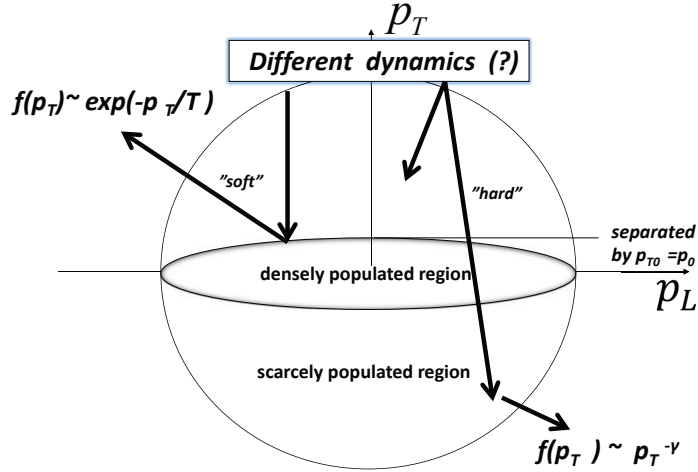}
\end{center}
\caption{ Schematic view of details of the phase space in
multiparticle production. Cf. text for details..} \label{FigPS}
\end{figure}

Multiparticle production processes, which will serve us as the
stage to present and discuss the ubiquitous presence of the
quasi-power law distributions, cover at present a $\sim 14$ orders
of magnitude span in the observed cross sections (when one
observes transverse momentum, $p_T$ distributions
\cite{CMS,ATLAS,ALICE}). It came as a surprise that this type of
data can be fitted by a one simple {\it quasi-power like} formula
\cite{CM,H}:
\begin{eqnarray}
  H(X) = C\cdot \left( 1 + \frac{X}{nX_0}\right)^{-n}
  \longrightarrow
  \left\{
 \begin{array}{l}
  \exp\left(-\frac{X}{X_0}\right)\quad \, \, \, {\rm for}\ X \to 0, \smallskip\\
  X^{-n}\qquad \qquad{\rm for}\ X \to \infty,
 \end{array}
 \right .
 \label{eq:H}
\end{eqnarray}
which smoothly combines pure power-like behavior in one part of
phase space with an exponential in another part. Before proceeding
further, a few words of explanation are in order. Fig. \ref{FigPS}
displays our playground, i.e., the phase space for particle
produced in high energy collision of, say protons, $p+p
\rightarrow N$ particles (mostly $\pi$ mesons). The initial energy
is $E_{in}$ and momenta of colliding particles are $\vec{p}_A$ and
$\vec{p}_B$. One usually works in center-of-mass system in which,
for $pp$ collisions, $\vec{p}_A = - \vec{p}_B $ and
$|\vec{p}_{A,B}| = P $). The momenta of produced secondaries are
decomposed in the longitudinal and perpendicular components (with
regards to the collision axis), $\vec{p} = \left[ p_L,
\vec{p}_T\right]$; their energies are $E = \left(\mu^2 +
p^2\right)^{1/2}$, where $\mu$ is mass of the produced secondary
and $p = |\vec{p}|$. Composition $\left(\mu^2 +
|\vec{p_T}|^2\right)^{1/2} = \mu_T$ is called transverse mass. It
is also customary to use variable $y = \ln \left[ \left(E +
p_L\right)/\left(E - p_L\right)\right]$ (rapidity) in which $p_L =
\mu_T \sinh y$ and $E = \mu_T \cosh y$. In what follows, we shall
be interested in the so called central rapidity region (with $ y
\simeq 0$, i.e., with $p_L \simeq 0$), and in distributions of
$p_T$ or $\mu_T$ only (i.e., in Eq. (\ref{eq:H}) $X=p_T$ or $X =
\mu_T$).

Different parts of phase space depicted schematically in Fig.
\ref{FigPS}. They are dominated by different collision dynamics.
It is customary to separate them artificially by some momentum
scale, transverse momentum parameter $p_0$, dividing transverse
phase space into a predominantly "hard" and predominantly "soft"
part. They distinguished themselves by the type of observed
spectra of secondaries produced. In the "hard" (scarcely
populated) region, with $p_T > p_0$, they are regarded as
essentially power-like, $F \left(p_T\right) \sim p_T^{-n}$, and
are usually associated with the hard scattering processes between
partons (constituents composing nucleons, quarks and gluons)
\cite{CYW}. In the "soft" (densely populated) region, with $p_T <
p_0$, the dominant distribution is exponential one, $F (p_T
)$$\sim$$\exp (-p_T/T)$. It is usually associated with the
thermodynamical description of the hadronizing system with $T$
playing the role of "temperature", with the fragmentation of a
flux tube with a transverse dimension, or with the production of
particles by the Schwinger mechanism \cite{CYW}. However, both
formulas can be unified in a single quasi-power like formula, Eq.
(\ref{eq:H}), interpolating smoothly between both regions. It
becomes power-like for high $p_T$ and exponential-like for low
$p_T$, as required. One way of introducing it is to start from the
very large values of $p_T$ where we have a pure, scale free, power
law. Decreasing now $p_T$ towards the demarcation value $p_0$ and
below it, the natural thing to avoid problems with unphysical
singularity for $p_T \rightarrow 0$ is to add to $p_T$ a constant
term, and choose it equal to $p_0$. In this way one gets Eq.
(\ref{eq:H}), which for small values of $X=p_T$ (and for
$X_0=p_0$) becomes exponential (Boltzmann-Gibbs - BG) distribution
with temperature $T= p_0$~~\footnote{Formula (\ref{eq:H}) is known
as the {\it QCD-based Hagedorn formula} \cite{H}. It was used for
the first time in the analysis of UA1 experimental data \cite{UA1}
and it became one of the standard phenomenological formulas for
$p_T$ data analysis.}.

This formula coincides with the so called Tsallis nonextensive
distribution \cite{Tsallis} for $n = 1/(q-1)$,
\begin{equation}
h_q(X)\, =\, C_q\cdot\left[ 1- (1-q)\frac{X}{X_0}
\right]^{\frac{1}{1-q}}\quad \stackrel{def}{=}\quad
C_q\cdot\exp_q\left(-\frac{X}{X_0}\right)\quad
 \stackrel{q \rightarrow 1}{\Longrightarrow}\quad C_1\cdot\exp
\left(-\frac{X}{X_0}\right).\label{eq:T}
\end{equation}
It has been widely used in many other branches of physics
\cite{Contemporary}. For our purposes, both formulas are
equivalent with $n$ = $1/(q-1)$ and $X_0=nT$, and we shall use
them interchangeably. Because Eq. (\ref{eq:T}) describes
nonextensive systems in statistical mechanics, the parameter $q$
is usually called the {\it nonextensivity parameter}. Eq. (2)
becomes the usual Boltzmann-Gibbs exponential distribution for $q$
$\to$ 1, with $T$ becoming the temperature. Both Eqs. (1) and (2)
have been widely used in data analysis (cite, for example,
\cite{CMS,ATLAS,ALICE,UA1,PHENIX,STAR}) and in the
phenomenological analysis of processes of multiparticle production
(cf., for example
\cite{BCM,Beck,RWW,WWrev,qWW,Wibig,B_et_all,JCleymans,ADeppman,Others,WalRaf}).

In the next Section \ref{sec:Examples} we present a review of a
number of examples of how such distributions occur (based on our
experience with applications of Tsallis statistics to
multiparticle production processes, but they have general
applicability). In Section \ref{sec:logosc} we present some
specific generalization of quasi-power distributions which allows
to account for a log-periodic oscillations in variable considered.
This phenomenon was so far known and observed in all other
situations resulting in power-like distributions. However, recent
multiparticle production data \cite{CMS,ATLAS,ALICE} seem to
indicate that, apparently, this phenomenon also starts to be
visible there and deserves attention. Section \ref{sec:Summary}
contains a summary.

\section{How to get a Tsallis distribution - some examples}
\label{sec:Examples}

In what follows we shall concentrate on Tsallis distributions, Eq.
(\ref{eq:T}), obtained from approaches not based on in
nonextensive thermodynamics\footnote{These are discussed in
\cite{JCleymans,ADeppman}; their theoretical justification is
presented in \cite{TherCons}.}. We start from examples of
constrained systems which lead to Tsallis distributions with $q <
1$. To get also $q > 1$ one has to allow for some intrinsic
fluctuations (or relaxing constraints). This will be shown next.

\subsection{All variables fixed} \label{sec:AllF}

In statistical physics, the simplest situation considered is when
a system is characterized by variables $U$ - energy, $T$-
temperature and $N$ - multiplicity. Usually one or two of them are
fixed and the rest fluctuates (either according to gamma
distribution, in case of $U$ or $T$, or according to Poisson
distribution in case of $N$, which are integers) \cite{WWcov}.
Only in the thermodynamic limit (i.e., for $N \rightarrow \infty$)
fluctuations take the form of Gaussian distributions usually
discussed in textbooks. In \cite{WWcov} we also discussed in
detail situations when all three variables fluctuate inducing some
correlations in the system.

However, if all variables are fixed we have distributions of the
type of
\begin{equation}
f(E) = \left(1 - \frac{E}{U}\right)^{N-2}\qquad {\rm with}\qquad q
= \frac{N-3}{N-2} < 1, \label{eq:ConstN}
\end{equation}
i.e., Tsallis distributions with $q < 1$. Interestingly enough,
such a distribution emerges also directly from the calculus of
probability for a situation known as {\it induced partition}
\cite{IndPart}. In short: $N-1$ randomly chosen independent points
$\left\{ U_{1},\dots, U_{N-1}\right\}$ break segment $(0,U)$ into
$N$ parts, the length of which is distributed according to Eq.
(\ref{eq:ConstN}). The length of the $k^{th}$ part corresponds to
the value of energy $E_k = U_{k+1}-U_k$ (for ordered $U_k$). One
could think of some analogy in physics to the case of random
breaks of string in $N-1$ points in the energy space. Notice that
the induced partition differs from {\it successive sampling} from
the uniform distribution, $E_k\in \left[ 0,U-E_1-E_2- \dots
-E_{k-1} \right]$, which results in $f(E) =1/E$~~\cite{ONEoverE}.

\subsection{Conditional probability} \label{sec:Cp}

To the category of constrained systems also belongs an example of
conditional probability. Consider a system of $n$ independent
points with energies $\left\{E_{i=1,\dots,N}\right\}$, each energy
distributed according to a Boltzmann distribution $g_i\left(
E_i\right)$ (i.e., their sum, $U = \sum_{i=1}^{N}E_i$, is then
distributed according to a gamma distribution $g_N(U)$):
\begin{equation}
g_i\left(E_i\right) = \frac{1}{\lambda}\exp\left(
-\frac{E_i}{\lambda} \right)\qquad{\rm and}\qquad g_N(U) =
\frac{1}{\lambda(N-1)}\left(\frac{U}{\lambda}\right)^{N-1} \exp
\left( - \frac{U}{\lambda}\right). \label{eq:BolGam}
\end{equation}
If the available energy is limited, $U = N\alpha= const$, the
resulting {\it conditional probability}
\begin{eqnarray}
f\left( E_i|U=N\alpha \right) = \frac{g_1\left( E_i \right)
g_{N-1}\left( N\alpha-E_i \right )}{g_N (N\alpha)} =
\frac{(N-1)}{N\alpha}\left(1 -
\frac{1}{N}\frac{E_i}{\alpha}\right)^{N-2} =
\frac{2-q}{\lambda}\left[ 1 - (1 - q) \frac{E_i}{\lambda}
\right]^{\frac{1}{1 - q}}, \label{eq:constraints}
\end{eqnarray}
becomes a Tsallis distribution with
\begin{equation}
q = \frac{N-3} {N-2} < 1\qquad {\rm and}\qquad \lambda =
\frac{\alpha N}{N-1}. \label{eq:c1}
\end{equation}

\subsection{Statistical physics considerations} \label{sec:SPc}

Both above results arise more formally  from {\it statistical
physics considerations} of isolated systems with energy $U =
const$ and with $\nu$ degrees of freedom ($\nu$ particles). Choose
a single degree of freedom with energy $E << U$ (i.e., the
remaining, or reservoir, energy is $E_r = U - E$). If this degree
of freedom is in a single, well defined, state then the number of
states of the whole system is $\Omega(U-E)$ and probability that
the energy of the chosen degree of freedom is $E$ is $P(E) \propto
\Omega(U-E)$. Expanding (slowly varying)
\begin{equation}
\ln \Omega(U-E) = \sum_{k=0}^{\infty} \frac{1}{k!}
\frac{\partial^{(k)} \ln \Omega}{\partial E_r^{(k)}}, \quad {\rm
with}\quad \beta = \frac{1}{k_B T} \stackrel{def}{=}
\frac{\partial \ln \Omega\left(E_r\right)}{\partial
E_r},\label{eq:deriv}
\end{equation}
around $U$, and keeping only the two first terms one gets
\begin{equation}
\ln P(E) \propto \ln \Omega(E) \propto - \beta E,\quad {\rm or}
\quad P(E) \propto \exp( - \beta E), \label{eq:BoltzmannD}
\end{equation}
i.e., a Boltzmann distribution (for which $q=1$). On the other
hand, one usually expects
\begin{equation}
\Omega\left(E_r\right) \propto
\left(\frac{E_r}{\nu}\right)^{\alpha_1 \nu - \alpha_2}
\label{eq:Omega}
\end{equation}
(where $\alpha_{1,2}$ are of the order of unity; we put $\alpha_1
= 1$ and, to account for diminishing the number of states in the
reservoir by one state, $\alpha_2 = 2$) \cite{Reif}. One can than
write that
\begin{equation}
\frac{\partial^k \beta}{\partial E_r^k}\, \propto\, (-1)^k k!
\frac{\nu - 2}{E^{k+1}_r}\, =\, (-1)^k k! \frac{\beta^{k-1}}{(\nu
- 2)^k}. \label{eq:FR}
\end{equation}
Because
$$ \ln(1+x) = \sum_{k=0}^{\infty}(-1)^k
\frac{x^{k+1}}{(k+1)}, $$ the full series for probability of
choosing energy $E$ can be written as:
\begin{eqnarray}
P(E) \propto \frac{\Omega(U-E}{\Omega(U)} = \exp\left[
\sum_{k=0}^{\infty}\frac{(-1)^k}{k+1}\frac{1}{(\nu - 2)^k}(- \beta
E)^{k+1}\right] = C\left(1 - \frac{1}{\nu - 2}\beta E\right)^{(\nu
- 2)} = \beta(2-q)[1 - (1-q)\beta E]^{\frac{1}{1-q}}.
\label{eq:statres}
\end{eqnarray}
For $q= 1 - 1/(\nu -2) \leq 1$ this result coincides with the
previous results from the induced partition and conditional
probability.

\subsection{Systems with fluctuating multiplicity $N$}
\label{sec:FN}

So far, we were getting Tsallis distributions with $q < 1$. To
obtain $q>1$ one has to relax the restrictions imposed on the
system considered; for example, by allowing for fluctuations of
one of the variables $U$, $N$ or $T$. Fluctuations of $T$, known
as {\it superstatistics}, were considered in \cite{SuperS,WW,BJ}.
Here we shall consider multiplicity $N$ fluctuating according to
some distribution $P(N)$. In this case, the resulting distribution
is
\begin{equation}
f(E) = \sum f_N(E) P(N), \label{eq:resdistr}
\end{equation}
where
\begin{equation}
f_N(E) = \left(1 - \frac{E}{U}\right)^N\qquad {\rm and}\qquad U =
\sum E = const, \label{eq:fixedN}
\end{equation}
is a distribution for fixed $N$ (to simplify notation we changed
$N-2$ in Eq. (\ref{eq:ConstN}) to $N$). The most characteristic
for our purposes distributions $P(N)$ are the, respectively, {\it
Binomial Distribution}, $P_{BD}$, {\it Poissonian Distribution},
$P_{PD}$ and {\it Negative Binomial Distributions}, $P_{NBD}$ (cf.
\cite{WWjets}):
\begin{eqnarray}
P_{BD}(N) &=& \frac{M!}{N!(M-N)!}
\left(\frac{<N>}{M}\right)^N\left(1 -
\frac{<N>}{M}\right)^{M-N}; \label{eq:BD}\\
P_{PD}(N)&=& \frac{<N>^N}{N!} e^{-\langle N\rangle}; \label{eq:PD}\\
P_{NBD}(N) &=& \frac{\Gamma(N+k)}{\Gamma(N+1)\Gamma(k)}\left(
\frac{<N>}{k}\right)^N \left( 1 + \frac{<N>}{k}\right)^{-k-N}.
\label{eq:NBD}
\end{eqnarray}
They lead, respectively, to Tsallis distributions with $q$ ranging
from $q < 1$ for the Binomial Distribution, $P_{BD}$,
Eq.(\ref{eq:BDf}), via $q=1$ Boltzmann distribution for Poissonian
Distribution, $P_{PD}$, Eq.(\ref{eq:PDf}), to $q > 1$ for the
Negative Binomial Distribution, $P_{NBD}$, Eq.(\ref{eq:NBDf}) (in
all cases $\beta = \langle N\rangle/U$):
\begin{eqnarray}
f_{BD}(E) &=& \left(1 - \frac{\beta
E}{M}\right)^M,\qquad q = 1 - \frac{1}{M}  < 1;  \label{eq:BDf}\\
f_{PD}(E) &=& \exp( - \beta E),\qquad\quad q = 1; \label{eq:PDf}\\
f_{NBD}(E) &=& \left( 1 + \frac{\beta E}{k}\right)^{-k}, \qquad q
= 1 + \frac{1}{k} > 1. \label{eq:NBDf}
\end{eqnarray}
In all three cases the physical meaning of the parameter $q$ is
the same: it measures the strength of multiplicity fluctuations,
\begin{equation}
q - 1 = \frac{Var(N)}{<N>^2} - \frac{1}{<N>}. \label{eq:FluctN}
\end{equation}
For BD one has $Var(N)/<N> < 1$, therefore for it $q <1$. For PD
$Var(N)/<N> = 1$, i.e., $q = 1$ as well. For NBD, where
$Var(N)/<N>~ > 1$, one has $q > 1$.

In the case of $q > 1$, i.e., for NDB, fluctuations of
multiplicity $N$ can be translated into fluctuations of
temperature $T$. This is possible because, as shown in
\cite{WWrev,NBDder}, NBD, which can be written also in the
following form,
$$P(N) = \frac{\Gamma(N+k)}{[\Gamma(N+1)\Gamma(k)]}\cdot\gamma^k(1+\gamma)^{-k-N},$$
can be obtained from the Poisson multiplicity distribution, $P(N)=
\frac{\bar{N}^N}{N!} e^{-{\bar{N}}},$ by fluctuating mean
multiplicity $\bar{N}$ using gamma distribution\footnote{We have
two types of average here: $\bar{X}$ means average value in a
given event whereas $<X>$ denotes averages over all events (or
ensembles).},
$$f(\bar{N}) = \frac{\gamma^k{\bar{N}}^{k-1}}{\Gamma(k)}\cdot e^{-\gamma \bar{N}}\quad
{\rm  with}\quad \gamma = \frac{k}{<\bar{N}>}.$$ Identifying
fluctuations of mean multiplicity $\bar{N}$ with fluctuations of
temperature $T$, one can express the above observation via
fluctuations of temperature. Namely, noticing that
$$\bar{\beta} = \frac{\bar{N}}{U},\qquad <\bar{N}> =
U<\bar{\beta}>\quad {\rm and}\quad \gamma =
\frac{k}{U<\bar{\beta}>},$$ one can rewrite the gamma distribution
for mean multiplicity, $f(\bar{N})$, as a gamma distribution of
mean inverse temperature $\vec{\beta}$,
\begin{eqnarray}
f\left( \bar{\beta}\right)\, =\,
\frac{k}{<\bar{\beta}>\Gamma(k)}\left(
\frac{k\bar{\beta}}{<\bar{\beta}>}\right)^{k-1}
\exp\left(-\frac{k\bar{\beta}}{<\bar{\beta}>}\right)\, \, =\, \,
\frac{ \left(\frac{1}{q-1}\frac{\bar{\beta}}{{<\bar{\beta}>}}
\right)^{\frac{1}{q - 1} -
1}}{(q-1)<\bar{\beta}>\Gamma\left(\frac{1}{q-1}\right)}
\exp\left(-\frac{1}{q-1}\frac{\bar{\beta}}{<\bar{\beta}>}\right).
\label{eq:NtoT}
\end{eqnarray}
And this is precisely the gamma distribution describing
temperature fluctuations, derived, used and investigated in {\it
superstatistics} \cite{WW,BJ,SuperS}. When convoluted with the
Boltzmann-Gibbs distribution, with $\beta$ as scale parameter, it
carries it into Tsallis distribution with parameter
\begin{equation}
q = 1 + \frac{Var(\bar{\beta})}{\langle
\bar{\beta}\rangle}\label{eq:FluctT}
\end{equation}
replacing previous Eq. (\ref{eq:FluctN}) and now denoting the
strength of temperature fluctuations.

\subsection{Preferential attachment} \label{sec:Sn}

So far, to get $ q > 1$ we were demanding the existence in the
system some form of intrinsic (i.e., nonstatistical) fluctuations.
However, the same effect can be obtained if the system exhibits
correlations of the preferential attachment type, corresponding to
the "rich-get-richer" phenomenon in networks
\cite{nets1,nets,nets2}, and if the scale parameter depends on the
variable under consideration. If
\begin{equation}
T\rightarrow T_0'(E) = T_0 + (q-1)E, \label{eq:pat}
\end{equation}
then the probability distribution function, $f(E)$, is given by an
equation the solution of which is a Tsallis distribution (again,
with $q > 1$):
\begin{equation}
\frac{df(E)}{dE} = - \frac{1}{T_0'(E)}f(E)\quad
\Longrightarrow\quad  f(E) = \frac{2-q}{T_0}\left[ 1 -
(1-q)\frac{E}{T_0}\right]^{\frac{1}{1-q}}. \label{eq:nets}
\end{equation}
For $T_0'(E) = T_0$ one gets again the usual exponential
distribution. This approach was also applied to an analysis of
multiparticle production processes in \cite{nets2}.

The "preferential attachment" can be also obtained from
superstatistics. As shown above, fluctuations of multiplicity $N$
are equivalent to the results of application of superstatistics,
where the convolution
\begin{equation}
f(E) = \int g(T) \exp\left( - \frac{E}{T}\right) dT \label{eq:SSt}
\end{equation}
becomes a Tsallis distribution, Eq. (\ref{eq:T}), for
\begin{equation}
g(T) = \frac{1}{\Gamma(n)T}\left(\frac{nT_0}{T}\right)^n
\exp\left( - \frac{nT_0}{T}\right). \label{eq:gamma}
\end{equation}
Differentiating Eq. (\ref{eq:SSt}) one obtains
\begin{equation}
\frac{df(E)}{dE} = - \frac{1}{T(E)} f(E)\qquad {\rm where}\quad
T(E) = T_0 + \frac{E}{n}. \label{eq:diffSSt}
\end{equation}
This is nothing else but the "preferential attachment" case, again
resulting in a Tsallis distribution, which for $T(E) = T_0$
becomes BG distribution, cf. Eq. (\ref{eq:nets})\footnote{This is
not the only place where such a form of $T(E)$ appears. For
example, in \cite{WalRaf} it occurs in a description of the
thermalization of quarks in a quark-gluon plasma by a collision
process treated within Fokker-Planck dynamics.}.

\subsection{Multiplicative noise}
\label{sec:Mn}

Tsallis distribution can also be obtained from {\it multiplicative
noise} \cite{WW,BJ} defined by the following Langevin equation
\cite{BJ},
\begin{equation}
\frac{dp}{dt} + \gamma(t)p = \xi(t). \label{eq:Le}
\end{equation}
Here $\gamma(t)$ and $\xi(t)$ denote stochastic processes
corresponding to, respectively, multiplicative and additive
noises. The resulting Fokker-Planck equation for distribution
function $f$,
\begin{equation}
\frac{\partial f}{\partial t} = - \frac{\partial \left( K_1
f\right)}{\partial p} + \frac{\partial^2 \left( K_2
f\right)}{\partial p^2}, \label{eq:FPe}
\end{equation}
where
\begin{equation}
K_1 = \langle \xi\rangle - \langle \gamma\rangle p\quad {\rm
and}\quad K_2 = Var(\xi) - 2 Cov(\xi, \gamma) p + Var(\gamma )
p^2, \label{eq:K1K2}
\end{equation}
has a stationary solution $f$, which satisfies
\begin{equation}
\frac{d \left( K_2 f\right)}{dp} = K_1 f. \label{eq:K1vK2}
\end{equation}
If there is no correlation between noises and no drift term due to
the additive noise, i.e., for $Cov(\xi,\gamma) = \langle
\xi\rangle = 0$ \cite{AT}, the solution of this equation is a
Tsallis distribution for $p^2$,
\begin{equation}
f(p) = \left[1 + (q - 1)\frac{p^2}{T}\right]^{\frac{q}{1-q}}~~{\rm
with}~~ T = \frac{2Var(\xi)}{\langle \gamma \rangle};~~q = 1 +
\frac{2Var(\gamma)}{\langle \gamma\rangle}. \label{eq:solutionpp}
\end{equation}
If we insist on a solution in the form of Eq. (\ref{eq:H}),
\begin{equation}
f(p) = \left[1 + \frac{p}{nT}\right]^n\qquad {\rm with}\qquad n =
\frac{1}{q-1}, \label{eq:singlep}
\end{equation}
then the condition to be satisfied has the form:
\begin{equation}
K_2(p) = \frac{nT+p}{n}\left[ K_1(p) - \frac{dK_1(p)}{dp}\right].
\label{eq:K1vK2p}
\end{equation}
One then gets a Tsallis distribution (\ref{eq:singlep}) but now
\begin{equation}
n = 2 + \frac{\langle \gamma\rangle}{Var(\gamma)}\quad{\rm
or}\quad q = 1 + \frac{Var(\gamma)}{\langle \gamma\rangle + 2
Var(\gamma)} \label{eq:n-q}
\end{equation}
and  $T$ becomes a $q$-dependent quantity (reminiscent of
effective temperature $T_{eff}$ as introduced by us in
\cite{WWrev}):
\begin{equation}
T(q) = (2-q)\left[ T_0 + (q-1)T_1\right]\quad{\rm with}~~
T_0=\frac{Cov(\xi,\gamma)}{\langle \gamma\rangle},~~T_1 =
\frac{\langle \xi\rangle}{2\langle \gamma\rangle}.
\label{eq:TeffN}
\end{equation}

\subsection{From Shannon entropy to Tsallis distribution}
\label{sec:TS}

As shown in \cite{TfromS}, a Tsallis distribution emerges in a
natural way from the usual Shannon entropy, $S$ (for some
probability density $f(x)$), by means of the usual MaxEnt
approach, if only one imposes the right constraint provided by
some function of $x$, $h(x)$:
\begin{equation}
S = - \int dx f(x)\ln[f(x)]\quad{\rm with~~constraint}\quad <h(x)>
= \int dx f(x)h(x) = const. \label{eq:StT}
\end{equation}
This approach contains the same information as that based on
Tsallis entropy. In fact, one can either use Tsallis entropy with
relatively simple constraints, or the Shannon entropy with rather
complicated ones (cf., for example, a list of possible
distributions one can get in this way \cite{GR}). One gets:
\begin{equation}
f(x) = \exp\left[ \lambda_0 + \lambda h(x)\right],
\label{eq:TfromS}
\end{equation}
with constants $\lambda_0$ and $\lambda$ calculated from the
normalization of $f(x)$ and from the constraint equation.  A
constraint
\begin{equation}
<z> = z_0 = \frac{q-1}{2-q}\qquad {\rm where}\qquad  z =
\ln\left[1 - (1 - q)\frac{E}{T_0}\right], \label{eq:constrT}
\end{equation}
results in a Tsallis distribution (remember that $f(z)dz=f(E)dE$),
\begin{equation}
f(z) = \frac{1}{z_0}\exp\left( -\frac{z}{z_0}\right)\quad
\Longrightarrow\quad  f(E) = \frac{1}{\left(1 +
z_0\right)T_0}\left( 1 + \frac{z_0}{1+z_0}\frac{E}{T_0}\right)^{ -
\frac{1 + z_0}{z_0}} =\quad
 \frac{2 - q}{T_0}\left[ 1 - (1 -
q)\frac{E}{T_0}\right]^{\frac{1}{1 - q}}. \label{eq:TfS}
\end{equation}
To obtain $T_0$, one has to assume the knowledge of $\langle
E\rangle$ (this would be the an only constraint in the case of BG
distribution but here it is additional condition to be accounted
for).

Although at the moment there is no clear understanding of the
physical meaning of the constraint (\ref{eq:constrT}) (except its
obvious usefulness in getting Eq. (\ref{eq:TfS})), it seems to be
a natural one from the point of view of the multiplicative noise
approach represented by Eq. (\ref{eq:Le}). That is because there
is a connection between the kind of noise occurring in Eq.
(\ref{eq:Le}) and the condition imposed in the MaxEnt approach
\cite{WW_AIP}. Namely, for processes described by an additive
noise, $dx/dt = \xi(t)$, the natural condition is that imposed on
the arithmetic mean, $<x>= c+\langle \xi\rangle t$, and it results
in exponential distributions. For the multiplicative noise, $dx/dt
= x\gamma(t)$, the natural condition is that imposed on the
geometric mean, $ <\ln x> = c+\langle \gamma\rangle t$, which
results in a power law distribution \cite{R}. It seems, therefore,
that condition (\ref{eq:constrT}) combines both possibilities and
leads to a quasi-power law Tsallis distribution combining both
types of behavior.

\subsection{Tsallis and QCD}
\label{sec:TQCD}

Recent high energy experiments from the Large Hadron Collider at
CERN (CMS \cite{CMS}, ATLAS \cite{ATLAS} and ALICE \cite{ALICE})
provided distributions of transverse momenta measured in the
previously unprecedented range $p_T \leq 180$ GeV. The measured
cross section then spans the range of $\sim 14$ orders of
magnitude. In \cite{qQCD} it was shown that all these data can be
successfully fitted by Tsallis distribution (\ref{eq:H}) or
(\ref{eq:T}), which is since then widely used in this case (cf.,
Fig. \ref{Fig1} as example). This caused question, how is it
possible because this is the usual domain reserved for the purely
perturbative QCD approach? In \cite{qQCD1} it was demonstrated
that:
\begin{itemize}
\item Starting from the pure QCD partonic picture od elementary
collisions ("hard" scatterings between quarks and gluons of
incoming protons proceeding with high momentum transfer) one gets
power index $n \simeq 4 - 4.5$. However one observes hadrons which
are formed from quarks and gluons by means of complicated
branching and fragmentation processes. It turns out that all these
processes can be parameterized in a relatively simple way and one
can easily reproduce $n \simeq 7-8$ as observed in the experiment
(depending on the energy of collision).

\item However, in this way one reproduces properly only the power
index $n$ (or $q$) and resulting distribution is of pure
power-like type, $\sim 1/p_T$, diverging for $p_T \rightarrow 0$
instead of being exponential there. To get a Tsallis distribution
one has to make the same phenomenological step as that proposed to
obtain Eq. (\ref{eq:H}): to replace $p_T \rightarrow p_{T0} + p_T$
in the denominator of $1/p_T$.  So far, the only rationale behind
this is that, in the QCD approach, large $p_T$ partons probe small
distances (with small cross sections). With diminishing of $p_T$,
these distances become larger (and cross sections are increasing)
and, eventually, they start to be of the order of the nucleon size
(actually it happens around $p_T \sim p_{T0}$). At this moment the
cross section should stop rising, i.e., it should not depend
anymore on the further decreasing of transverse momentum $p_T$.
This can be modelled by introducing the constant term as above.
Effectively one then has $\left( 1/p_T\right)^{-n} \rightarrow
\left[ p_{T0}\cdot \left( 1 + p_T/p_{T0} \right)\right]^{-n}$. The
scale parameter $p_{T0}$ can then be identified with that in Fig.
\ref{FigPS}. In a Tsallis fit we use one formula for the whole
phase space with $p_{T0}$ becoming $T$ in the exponent for small
transverse momenta, and scale parameter for large $p_T$. Usually
one uses an exponential formula for $p_T < p_{T0}$ and power for
$p_T > p_{T0}$ and $p_{T0}$ separates the two parts of phase
space.

\end{itemize}

\section{Log-periodic oscillations}
\label{sec:LPO}

\begin{figure}[h]
\begin{center}
\includegraphics[width=9.5cm]{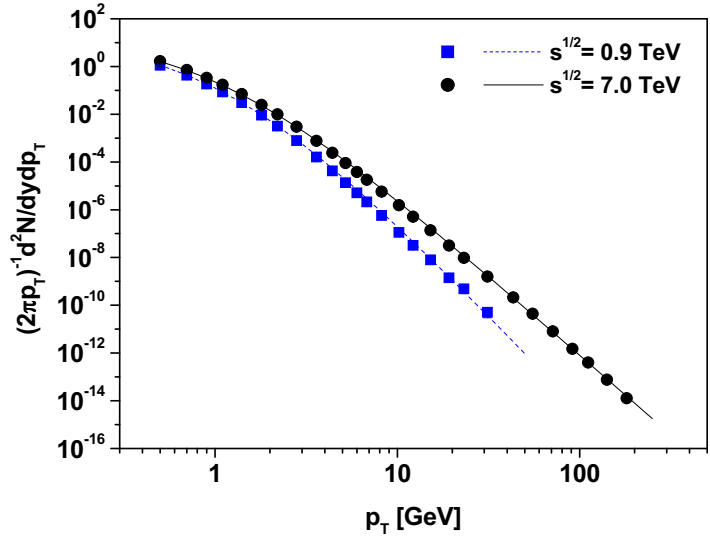}
\end{center}
\caption{(Color online) Fit to large $p_T$ data for $pp$
collisions at $0.9$ and $7$ TeV from CMS experiment using
distribution (\ref{eq:H}) \cite{CMS}. Parameters used are,
respectively, $(T=0.135,~m=8)$ and $(T=0.145,~m=6.7)$.}
\label{Fig1}
\end{figure}

\begin{figure}[h]
\begin{center}
\hspace{-3mm}
\includegraphics[width=8.cm]{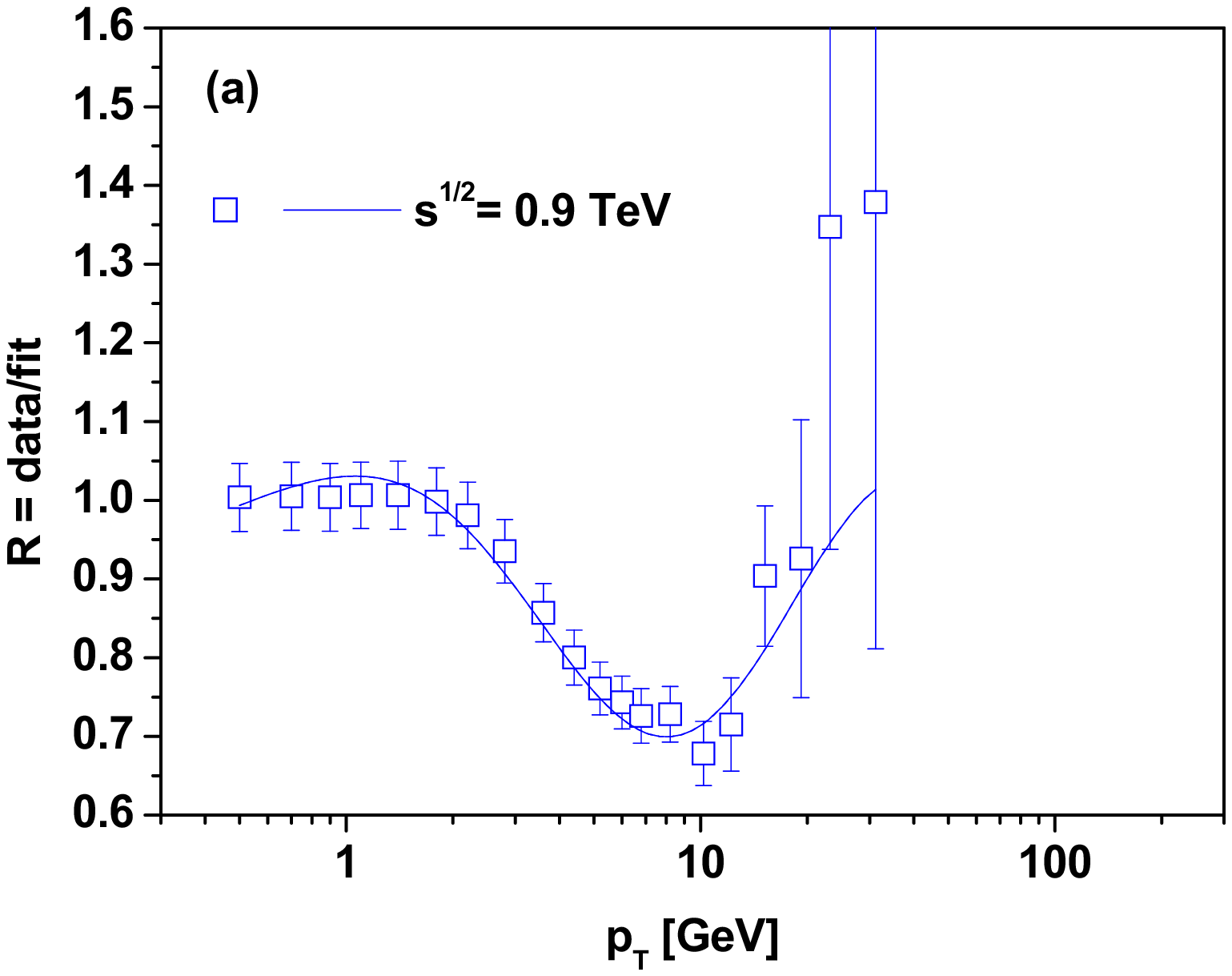}\hfill
\includegraphics[width=8.cm]{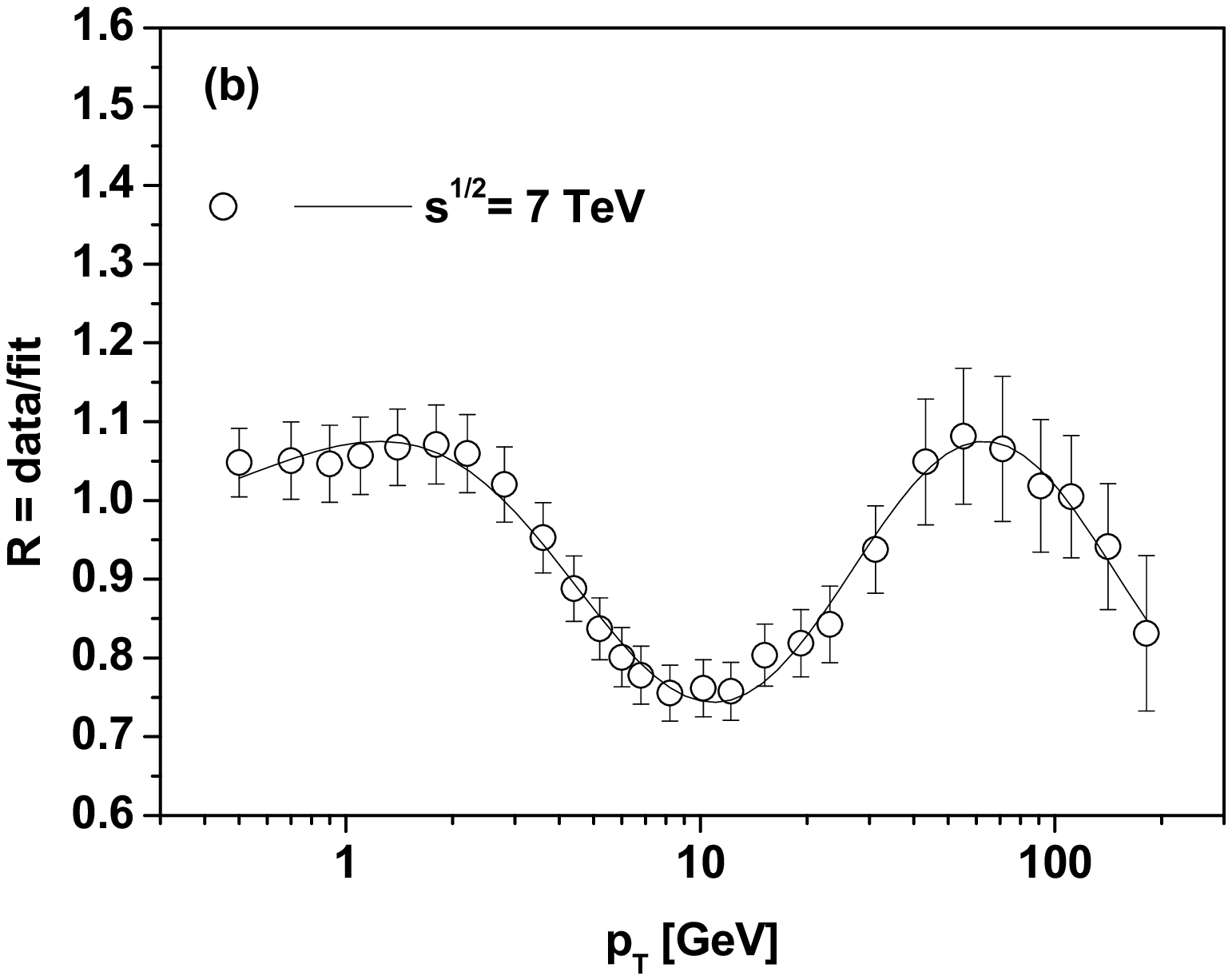}
\end{center}
\caption{(Color online) Fit to $p_T$ dependence of data/fit ratio
for results from Fig. \ref{Fig1}. Parameters of function $R$ used
(\ref{eq:Factor}) here are, respectively: $a=0.865$, $c=2.1$ for
$0.9$ TeV and $a=0.909$, $c=1.86$ for $7$ TeV, whereas for both
energies $b=0.166$, $d=0.948$ and $f=-1.462$.}
\label{Fig2}\vspace{-2mm}
\end{figure}

So far, we have presented possible derivations and applications of
relatively simple form of Tsallis distribution with Fig.
\ref{Fig1} as an example of its apparent success in fitting even
the most demanding data so far. However, closer inspection of Fig.
\ref{Fig1} shows that ratio of data/fit, usually used to estimate
the quality of fits, is not flat but shows some kind of clearly
visible oscillations of {\it log-periodic} character, cf. Fig.
\ref{Fig2}. The first conjunction in such a case is that
parameters used were not chosen in an optimal way. However, it
turns out that these oscillations cannot be eliminated by any
suitable changes of parameters $(q,T)$ or $(m,T)$ in Eqs.
(\ref{eq:T}) or (\ref{eq:H}), respectively. Here we shall
concentrate only on data from the CMS experiment \cite{CMS}, data
from ATLAS \cite{ATLAS} lead to identical conclusions. One also
has to realize that to really see these oscillations one needs a
rather large domain in $p_T$. Therefore, albeit similar effects
can also be seen at lower energies, they are not so pronounced as
here and will not be discussed here. Assuming that this is not an
experimental artifact one has to admit that it tells us that the
Tsallis formula used is too simple. There still remains something
hidden in data which has, so far, avoided to be disclosed, and
which can signal some genuine dynamical effect which is worth
been investigating in more detail.

Rather than look for another distribution we shall keep to Tsallis
distribution and attempt to improve it accordingly to account for
effects of these {\it log-periodic oscillations} observed in data.
Because we have two parameters here, power index $n$ and scale
("temperature") $T$, the natural approach is to modify one of
them. Because data, which we shall analyze below, are presented at
midrapidity (i.e., for $y \sim 0$ and longitudinal momentum $p_L
\sim 0$) and for large transverse momenta, $p_T >> \mu $, the
energy $E$ of a produced particle is essentially equal to its
transverse momentum $p_T$, which we shall use in what follows.

\subsection{Quasi-power laws with complex power indices}
\label{sec:logosc}

First notice that log-periodic oscillations are ubiquitous in
systems described in general by power distributions
\cite{LPO-examples}. Usually they suggest existence of some
scale-invariant hierarchical fine-structure in the system and
indicate its possible multifractality \cite{Scaling}. In the
context of nonextensive statistical mechanics log-periodic
oscillations have been first observed while analyzing the
convergence to the critical attractor of dissipative maps
\cite{LogMaps} and restricted random walks \cite{RRW}. In the case
of pure power like distributions the only parameter to manipulated
was the power index. It was then natural to modify this parameter
by allowing it to be complex \cite{LPO-examples,Scaling}.

For quasi-power Tsallis distributions this idea was first
investigated in detail by us in \cite{cqWW}. Here we shall apply
it to multiparticle data as mentioned before. The complex power
index results in effective {\it dressing} the original
distribution by multiplying it by a log-oscillating function,
usually taken in the form of:
\begin{equation}
R(E)= a + b\cos\left[ c\ln(E + d) + f\right]. \label{eq:Factor}
\end{equation}

In \cite{cqWW} we derived such a factor for a Tsallis
distribution. For completeness, we recapitulate the main points of
this derivation. In general, if one deals with a scale invariant
function, $O(x)$, i.e., if
\begin{equation}
O(\lambda x) = \mu O(x),\label{eq:SI}
\end{equation}
then it must have power law behavior,
\begin{equation}
O(x) = Cx^{-m}\quad {\rm with}\quad m = - \frac{\ln \mu}{\ln
\lambda}. \label{eq:PowerLaw}
\end{equation}
It means that, in general, one can write that ($k$ is an arbitrary
integer)
\begin{equation}
\mu \lambda^{m} = 1 = e^{i2\pi k}\quad \Longrightarrow\quad m = -
\frac{\ln \mu}{\ln \lambda} + i \frac{2\pi k}{\ln \lambda}.
\label{eq:cq}
\end{equation}
One must now find whether the Tsallis distribution has a similar
property and under what conditions. To this end we start from
differential $df(E)/dE$ of a Tsallis distribution $f(E)$ with
power index $n$, cf. Eq. (\ref{eq:diffSSt}), and write it for
finite differences \cite{cqWW},
\begin{equation}
\delta E = \alpha (nT + E), \label{eq:dE}
\end{equation}
where $\alpha n < 1$ is a new parameter. This leads to the
following scale invariant relation,
\begin{equation}
g[(1 + \alpha )x] = (1 - \alpha n)g(x) \label{eq:scaling}
\end{equation}
where
\begin{equation}
x = 1 + \frac{E}{nT}. \label{eq:Xvar}
\end{equation}
This means then that, in general, one can write Eq. (\ref{eq:H})
in the form:
\begin{equation}
 g(x) = x^{-m_k},\qquad\quad m_k = - \frac{\ln ( 1 - \alpha n)}{\ln (1 +
 \alpha)} + ik \frac{2\pi}{\ln(1 + \alpha)}. \label{eq:solution}
\end{equation}
The power index in Eq. (\ref{eq:solution}) (and therefore also in
Eq. (\ref{eq:H})) becomes a complex number, its imaginary part is
signaling a hierarchy of scales leading to the log-periodic
oscillations.

If we limit ourselves to $k=0$, one recovers the usual real power
law solution and $m_0$ corresponds to fully continuous scale
invariance. However, in this case the power law exponent $m_0$
still depends on $\alpha$ and increases with it roughly as
\begin{equation}
m_0 \simeq n + \frac{n}{2} (n+1)\alpha + \frac{n}{12}\left(4n^2 +
3n -1\right)\alpha^2 + \frac{n}{24}\left( 6n^3 + 4n^2 - n
+1\right)\alpha^3 + \dots.\label{eq:m0}
\end{equation}
The usual Tsallis distribution is recovered only in the limit
$\alpha \rightarrow 0$.

In general one has
\begin{figure}[t]
\begin{center}
\includegraphics[width=8.cm]{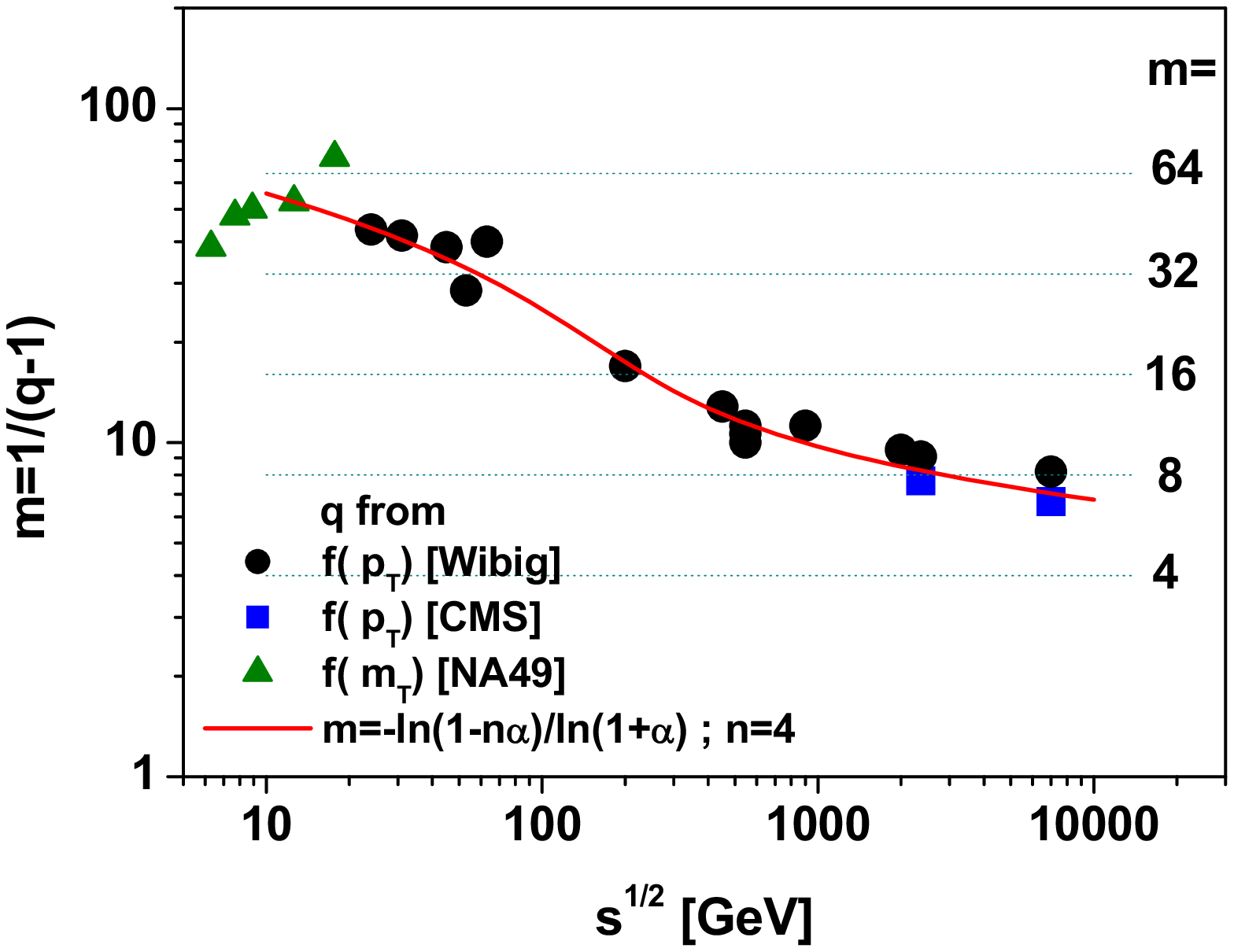}\hfill
\includegraphics[width=8.cm]{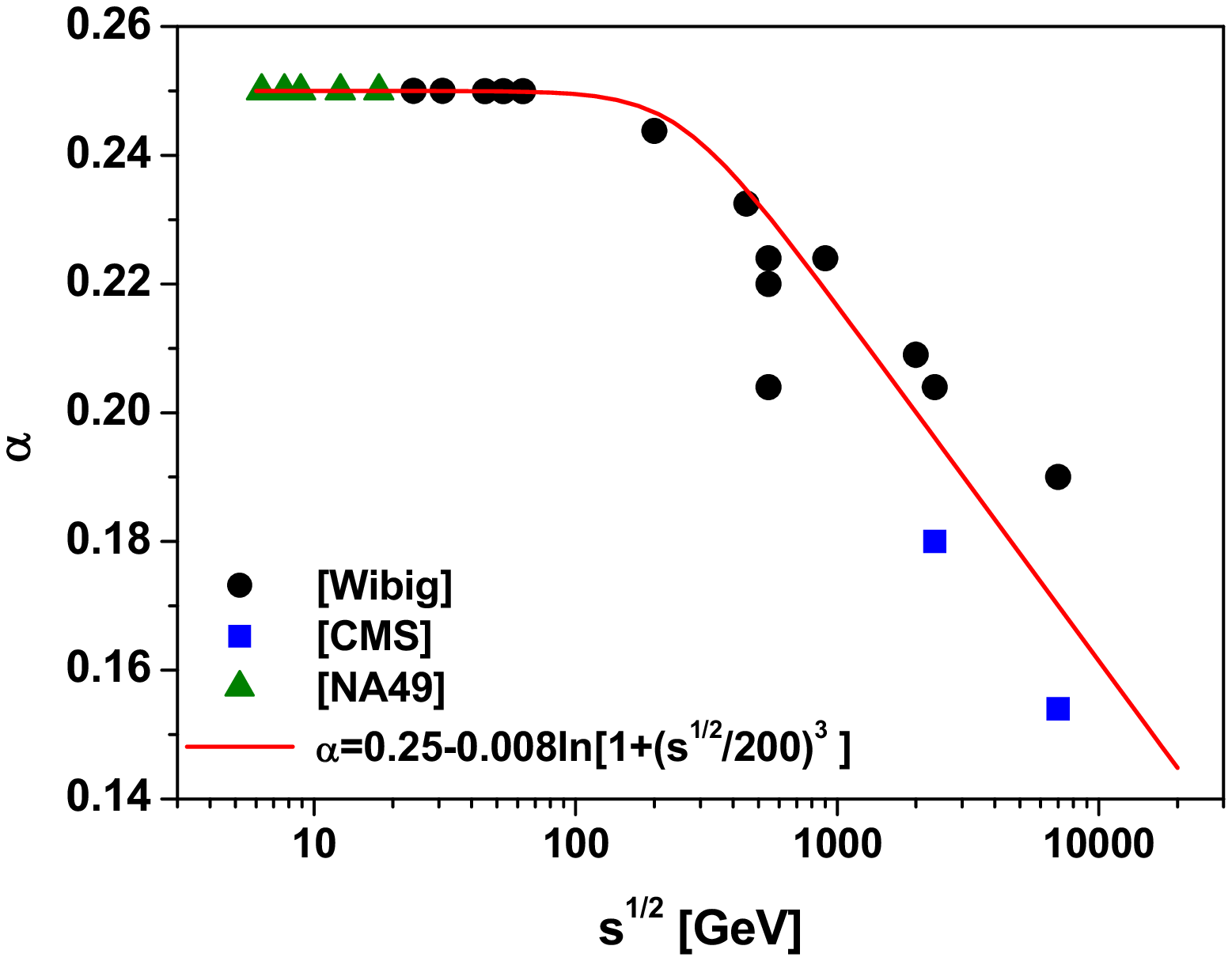}
\end{center}
\caption{(Color online)  Left panel: The energy dependence of $m =
m_0$ deduced from CMS \cite{CMS} and NA49 \cite{NA49} data and
from a compilation [Wibig] \cite{Wibig}. Right panel: The energy
dependence of parameter $\alpha$ present in $m_0$ plotted in left
panel.} \label{Fig3}
\end{figure}
\begin{eqnarray}
g(x) = \sum_{k=0}w_k\cdot {\rm Re}\left( x^{-m_k}\right) = x^{-
{\rm Re}\left( m_k\right)}\sum_{k=0}w_k\cdot \cos\left[ {\rm
Im}\left(m_k\right) \ln(x) \right]. \label{eq:fin}
\end{eqnarray}
This is a general form of a Tsallis distribution for complex
values of the nonextensivity parameter $q$. It consists of the
usual Tsallis form (albeit with a modified power exponent) and a
dressing factor which has the form of a sum of log-oscillating
components, numbered by parameter $k$. Because we do not know {\it
a priori} the details of dynamics of processes under consideration
(i.e., we do not known the weights $w_k$), in what follows we only
use $k=0$ and $k=1$ terms. We obtain approximately,
\begin{eqnarray}
g(E) \simeq \left( 1 + \frac{E}{nT}\right)^{-m_0}\cdot \left\{ w_0
+ w_1\cos\left[ \frac{2\pi}{\ln (1 + \alpha)} \ln \left( 1 +
\frac{E}{nT}\right)\right]\right\}. \label{eq:approx}
\end{eqnarray}
In this case one could expect that parameters in general
modulating factor $R$ in Eq. (\ref{eq:Factor}) could be identified
as follows:
\begin{equation}
a = w_0,\quad b = w_1,\quad c=\frac{2\pi}{\ln (1 + \alpha)},\quad
d = nT,\quad f = - c\cdot \ln(nT). \label{eq:parameters}
\end{equation}
Comparison of the fit parameters of the oscillating term  $R$ in
Eq. ({\ref{eq:Factor}) with Eq. (\ref{eq:solution}) clearly shows
that the observed frequency, here given by the parameter $c$, is
more than an order of magnitude smaller than the expected value
equal to $2\pi/\ln (1 + \alpha)$ for any reasonable value of
$\alpha$. To explain this, notice that in our formalism leading to
Eq. (\ref{eq:approx})) only one evolution step is assumed, whereas
in reality we have a whole hierarchy of $\kappa$ evolutions. This
results (cf. \cite{cqWW}) in the scale parameter $c$ being
$\kappa$ times smaller than in (\ref{eq:approx}),
\begin{equation}
c = \frac{2 \pi}{\kappa \ln (1 + \alpha)}. \label{eq:kappa}
\end{equation}
Experimental data indicate that $\kappa \simeq 22$  (for $\alpha
\simeq 0.15$ and $c \simeq 2$ ).

From Eq.(\ref{eq:solution}) we see that $m_0 > n$. This suggests
the following explanation of the difference seen between
prediction from theory and the experimental data: the measurements
in which log-periodic oscillations appear underestimate the true
value that follows from the underlying dynamics which leads to the
smooth Tsallis distribution. As an example consider the $m_0$
dependence on $\alpha$, assuming the initial slope $n = 4$ (this
is the value of $n$ expected from the pure QCD considerations for
partonic interactions \cite{qQCD1}). The energy behavior of the
power index $m_0$ in the Tsallis part is shown in the left panel
of Fig. \ref{Fig3}, whereas the energy dependence of the parameter
$\alpha$ contained in $m_0$ is shown in the right panel of Fig.
\ref{Fig3}.

\subsection{Quasi-power laws with log-periodic scale parameter $T$}
\label{sec:LPT}

The phenomenon of log-periodic oscillations observed in data  can
also be explained in a different way. We can keep the
nonextensivity parameter $q$ real (as in the original Tsallis
distribution) but allow the scale parameter $T$ to oscillate in a
specific way, as displayed in Fig. \ref{Fig5}. These oscillations
can be fitted by a formula similar to Eq. (\ref{eq:Factor}), with
generally energy dependent fit parameters $(\bar{a}, \bar{b},
\bar{c}, \bar{d}, \bar{f})$):
\begin{equation}
T = \bar{a} + \bar{b}\sin\left[ \bar{c} \left( \ln(E +
\bar{d}\right) + \bar{f} \right]. \label{eq:TT}
\end{equation}
\begin{figure}[h]
\begin{center}
\includegraphics[width=10cm]{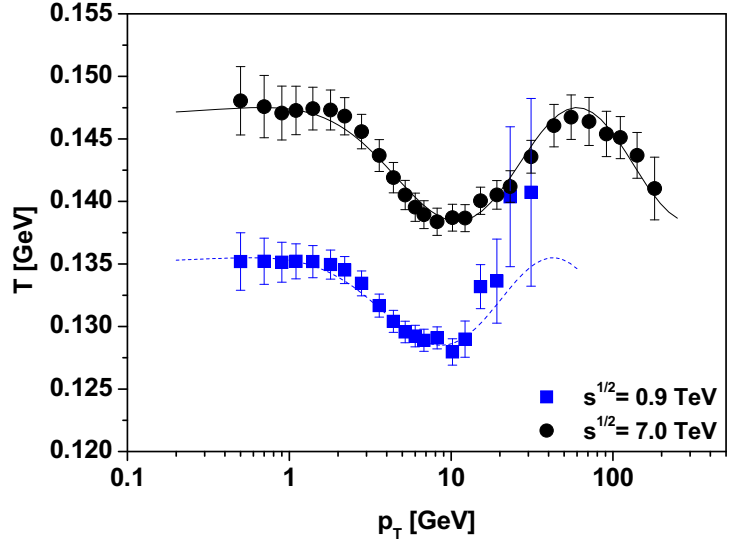}
\end{center}
\caption{(Color online) The $T=T\left( p_T\right)$ for Eq.
(\ref{eq:T}) for which $R=1$. Parameters used are: $\bar{a} =
0.132,~\bar{b} = 0.0035,~\bar{c} = 2.2,~\bar{d} = 2.0,~\bar{f} =
-0.5$ for $0.9$ TeV and $\bar{a} = 0.143,~\bar{b} =
0.0045,~\bar{c} = 2.0,~\bar{d} = 2.0,~\bar{f} = -0.4$ for $7$
TeV.} \label{Fig5}
\end{figure}

To get such behavior we start from the well known stochastic
equation for the temperature evolution \cite{Kampen}, which in the
Langevin formulation (allowing for an energy dependent noise,
$\xi(t,E)$) has the form:
\begin{equation}
\frac{dT}{dt} + \frac{1}{\tau} T + \xi(t,E) T = \Phi,
\label{eq:St1}
\end{equation}
$\tau$ is relaxation time which, for a while, we keep constant.
For the time dependent $E=E(t)$ it reads:
\begin{equation}
\frac{dT}{dE}\frac{dE}{dt} + \frac{1}{\tau} T + \xi(t,E) T = \Phi.
\label{eq:St2}
\end{equation}
In the scenario of {\it preferential attachment} known from the
growth of networks, cf. Section \ref{sec:Sn}, evolution equations
as given by Eqs. (\ref{eq:nets}) and (\ref{eq:diffSSt}) can be
derived from master equation $df(E,t)/dt = - f(E,t)$ for the
growth of network $dE/dt$ given by \cite{nets1}
\begin{equation}
\frac{dE}{dt} = \frac{E}{n} + T \label{eq:St3}
\end{equation}
($n$ coincides with power index in Eq. (\ref{eq:H})). This will
therefore be the equation we shall use in what
follows\footnote{Notice that in the usually multiplicative noise
scenario described by $\gamma(t)$, not discussed here, one has
$\frac{dE}{dt} = \gamma(t) E + \xi (t)$.}. With it one can write
Eq. (\ref{eq:St2}) as
\begin{equation}
\left(\frac{E}{n} + T\right)\frac{dT}{dE} + \frac{1}{\tau} T +
\xi(t,E) T = \Phi .\label{eq:St4}
\end{equation}
This can be subsequently transformed to
\begin{equation}
\left( \frac{1}{n} + T e^{-\ln E}\right)\frac{dT}{d(\ln E)} +
\frac{1}{\tau} T + \xi(t,E) T = \Phi \label{eq:eq:St4a}
\end{equation}
and, after differentiating, to
\begin{eqnarray}
\left( \frac{1}{n} + T e^{-\ln E}\right)\frac{d^2T}{d(\ln E)^2} +
\left[\frac{dT}{d(\ln E)}\right]^2 e^{-\ln E} - \left[  T e^{-\ln
E} - \frac{1}{\tau} - \xi(t,E)\right]\frac{dT}{d(\ln E)} +
T\frac{d\xi(t,E)}{d(\ln E)} = 0. \label{eq:St4b}
\end{eqnarray}
For large $E$ (i.e., neglecting terms $\propto 1/E$) one obtains
the following equation for $T$:
\begin{equation}
\frac{1}{n}\frac{d^2T}{d(\ln E)^2} + \left[ \frac{1}{\tau} +
\xi(t,E) \right]\frac{dT}{d(\ln E)} + T\frac{d\xi(t,E)}{d\ln E)} =
0.\label{eq:St5}
\end{equation}
To proceed further one has to specify the energy dependence of the
noise $\xi(t,E)$. We assume that it increases logarithmically with
energy in the following way,
\begin{equation}
\xi(t,E) = \xi_0(t) + \frac{\omega^2}{n}\ln E \label{eq:St6}
\end{equation}
(where $n$ is, again, power index from Eq. (\ref{eq:H}) and
$\omega$ is a new parameter). For this choice of noise Eq.
(\ref{eq:St5}) is just an equation for the damped hadronic
oscillator and has a solution in the form of a log-periodic
oscillation of temperature with frequency $\omega$:
\begin{equation}
T = C \exp\left\{ - n\cdot\left[\frac{1}{2\tau} +
\frac{\xi(t,E)}{2}\right]\ln E\right\}\cdot \sin(\omega\ln E +
\phi). \label{eq:St7}
\end{equation}
The phase shift parameter $\phi$ depends on the unknown initial
conditions and is therefore an additional fitting parameter.
Averaging the noise fluctuations over time $t$ and taking into
account that the  noise term cannot on average change the
temperature (cf. Eq. (\ref{eq:St1}) in which $\langle dT/dt\rangle
= 0$ for $\Phi = 0$), i.e., that
\begin{equation}
\frac{1}{\tau} + \langle \xi(t,E)\rangle = 0, \label{eq:St8}
\end{equation}
we have
\begin{equation}
T = \bar{a} + \frac{b'}{n}\sin( \omega \ln E + \phi).
\label{eq:St9}
\end{equation}
The amplitude of oscillations, $b'/n$, comes from the assumed
behavior of the noise as given in Eq. (\ref{eq:St6}). Notice that
for large $n$, the energy dependence of the noise disappears. It
means that, because, in general, $n$ decreases with energy
\cite{qQCD1}, one can expect only negligible oscillations for
lower energies but increase with the energy.

This should now be compared with the parametrization of $T\left(
p_T\right)$ given by Eq. (\ref{eq:TT}) and used to fit data in
Fig.\ref{Fig5}. Looking at parameters we can see that only a small
amount of $T$ (of the order of $\bar{b}/\bar{a} \sim 3\%$) comes
from the stochastic process with energy dependent noise, whereas
the main contribution emerges from the usual energy-independent
Gaussian white noise.

The above oscillating $T$ needed to fit the log-periodic
oscillations seen in data can be obtained in yet another way. So
far we were assuming that the noise $\xi(t,E)$ has the form of Eq.
(\ref{eq:St6}) and, at the same time, we were keeping the
relaxation time $\tau$ constant. In fact, we could equivalently
assume the energy $E$ independent white noise, $\xi(t,E) =
\xi_0(t)$, but allow for the energy dependent relaxation time.
Assuming it in the form,
\begin{equation}
\tau = \tau(E)  = \frac{n\tau_0}{n + \omega^2 \ln E},
\label{ew:tauE}
\end{equation}
results in the following time evolution of the temperature,
\begin{equation}
T(t) = \langle T\rangle + [T(t=0) - \langle T\rangle]
E^{-t\omega^2/n} \exp\left(-\frac{t}{\tau_0}\right).
\label{eq:Ttau}
\end{equation}
It gradually approaches its equilibrium value, $\langle T
\rangle$, and reaches it sooner for higher energies.

\subsubsection{Log-periodic oscillations: summary}
\label{sec:LPOSum}

To summarize, we have presented two possible mechanisms which
could result in the log-periodic oscillations apparently present
in data for transverse momentum distributions observed in LHC
experiments. In both cases one uses a Tsallis formula (either in
the form of Eq. (\ref{eq:H}) or Eq. (\ref{eq:T})), with main
parameters $m$ - the scaling exponent (or nonextensivity $q = 1 +
1/m$) and $T$ - the scale parameter (temperature).
\begin{itemize}
\item  In the first approach, our Tsallis distribution is
decorated by an oscillating factor. This is done by changing in
Eq. (\ref{eq:diffSSt}) differentials by finite differences,  $dE
\rightarrow \delta E = \alpha (nT + E)$, where the new parameter
$\alpha < 1/n$ regulates the smallness of $\delta E$. As results,
we get for $x = 1 + E/(nT)$ the scale invariant relation, $g[(1 +
\alpha )x] = (1 - \alpha n)g(x)$. This, in turn, means that power
index $m$ (and also nonextensivity parameter $q$) becomes a
complex number, of which the imaginary part describes a hierarchy
of scales leading to the log-periodic oscillations. The scale
parameter $T$ remains unaltered.

It should be mentioned at this point that complex $q$ inevitably
also means complex heat capacity $C = 1/(1 - q)$ (c.f.,
\cite{Campisi,qWW,BiroC}). Such complex (frequency dependent) heat
capacities (meaning relaxing temperature) are widely known and
investigated, see \cite{HC}.

\item In the second approach, it is the other way around, i.e.,
whereas $m = n$ remains untouched, the scale parameter $T$ is now
oscillating. From Eq. (\ref{eq:St9}) one can see that
$T=T(n=1/(1-q),E)$ and as a function of nonextensivity parameter
$q$ it continues our previous efforts to introduce an effective
temperature into the Tsallis distribution, $T_{eff} = T(q)$
\cite{WWrev,WWTout,WW_AIP} (but here in a much more general form).
The two possible mechanisms resulting in such $T$ were outlined:
$(i)$ - the energy dependent noise connected with the constant
relaxation time, or else, $(ii)$ - the energy independent white
noise, but with energy dependent relaxation time.
\end{itemize}

At the present level of investigation, we are not able to indicate
which of the two possible mechanisms presented here (complex $q$
or oscillating $T$) and resulting in log-periodic oscillations is
preferred. This would demand more detailed studies on the possible
connections with dynamical pictures. For example, as discussed
long time ago by studying apparently similar effects in some
exclusive reactions using the QCD Coulomb phase shift idea
\cite{Pire}. The occurrence of some kind of complex power
exponents was noticed there as well, albeit on completely
different grounds than in our case. A possible link with our
present analysis would be very interesting but would demand an
involved and thorough analysis.

\section{Summary and conclusions}
\label{sec:Summary}

We present examples of a possible mechanisms resulting in the
quasi-power distributions exemplified by Tsallis distribution Eqs.
(\ref{eq:H}) and (\ref{eq:T}). Our presentation is limited to
approaches not derived from nonextensive thermodynamic connections
of this distribution \cite{JCleymans,ADeppman,TherCons}.

It was shown that statistical physics consideration, as well as
"induced partition process", results in a Tsallis distribution
with $q = (N - 3)/(N - 2) < 1$, Eq. (\ref{eq:ConstN}). To get $q
> 1$ one has to allow for fluctuations of the multiplicity $N$. They
modify the parameter $q$ which is now $q = 1 + Var(N)/\langle
N\rangle^2 - 1/\langle N\rangle$, cf. Eq. (\ref{eq:FluctN}). The
conditional probability for BG distribution results again in Eq.
(\ref{eq:ConstN}).

We proposed and discussed two possible mechanisms which would
allow quasi-power law distributions to which Tsallis distribution
belongs, to describe data showing a log-periodic "decoration" of
simple power law distributions. One is a generalization of Tsallis
distribution to real power $n$ (it can be regarded as
generalization of such well known distributions as Snadecor
distribution (with $n = (\nu + 2)/2 $ with integer $\nu$, for $\nu
\rightarrow \infty$ it becomes an exponential distribution), can
be extended to complex nonextensivity parameter). The other is
introducing a specific, log-periodic oscillating scale parameter,
effective temperature $T_{eff}$, generalizing our previous results
in this field presented and used in \cite{WWrev,WW_AIP,WWTout}.

Different derivations of Tsalis distributions turn out to be, in a
sense, equivalent. Fluctuations of multiplicity $N$ are equivalent
to results fluctuations of $T$ (which is the basis of of
superstatistics); on the other hand, from a superstatistics
formula one can get the preferential attachment, cf. Sections
\ref{sec:FN} and \ref{sec:Sn}.

\section*{Acknowledgements}

This research  was supported in part by the National Science
Center (NCN) under contract DEC-2013/09/B/ST2/02897. We would like
to warmly thank Dr Eryk Infeld for reading this manuscript.

\end{document}